# Right Scaling for Right Pricing:
# A Case Study on Total Cost of Ownership
# Measurement for Cloud Migration

Pierangelo Rosati[1], Frank Fowley[1], Claus Pahl[2], Davide Taibi[3],
and Theo Lynn[1]

[1] Irish Centre for Cloud Computing and Commerce,
Dublin City University, Dublin, Ireland
`{pierangelo.rosati,frank.fowley,theo.lynn}@dcu.ie`
[2] Faculty of Computer Science, Free University of Bozen-Bolzano,
Bolzano, Italy
`claus.pahl@unibz.it`
[3] Laboratory of Pervasive Computing, Tampere University of Technology,
Tampere, Finland
`davide.taibi@tut.fi`

**Abstract.** Cloud computing promises traditional enterprises and independent software vendors a myriad of advantages over on-premise installations including cost, operational and organizational efficiencies. The decision to migrate software configured for on-premise delivery to the cloud requires careful technical consideration and planning. In this chapter, we discuss the impact of right-scaling on the cost modelling for migration decision making and price setting of software for commercial resale. An integrated process is presented for measuring total cost of ownership, taking in to account IaaS/PaaS resource consumption based on forecast SaaS usage levels. The process is illustrated with a real world case study.

## 1 Introduction

Cloud computing is increasingly the computing paradigm of choice for enterprises worldwide. Cloud computing is particularly attractive from a business perspective since it requires lower upfront capital expenditure, and improves operational and organizational efficiencies and agility [4, 9, 39, 45]. Similarly, from a technical perspective, the benefits of the cloud are well documented including on-demand and self-service capabilities, resource pooling and rapid elasticity [4]. However, the success of cloud computing investments highly depends on accurate and efficient decision making; the implications of investment decisions need to be quantifiable to allow a comparison of alternatives, both from the consumer's and from the vendor's perspective [27].

Cloud computing adoption may generate significant challenges particularly for software producers (SPs) offering a Software-as-a-Service (SaaS) model. SPs typically migrate their software to a third-party platform (Infrastructure-as-a-Service – IaaS – or Platform-as-a-Service – PaaS) and their customers access it from this new multi-tenant architecture. In a cloud environment both SPs and their customers are typically charged on a pay-per-use or subscription basis. Furthermore, SPs do not have control of customers' service usage; in such a context, it is crucial for SPs to identify the right architectural configuration to meet service level agreement (SLA) obligations at the minimum cost. Being charged on a per-use basis also represents a radical change in the producers' cost and revenue models and introduces additional uncertainty in cash flow forecasting [15]. Furthermore, the actual cost of the migration process might be substantial for SPs and for their legacy customers, while nonexistent for cloud-native SPs. According to the Cloud Native Computing Foundation, modern cloud-native systems have the following properties:

- Container-packaged;
- Dynamically managed by a central orchestrating process;
- Microservice-oriented.

Cloud-native architectures have technical advantages in terms of isolation and reusability, thus reducing cost for maintenance and operations. PaaS clouds with their recent support for containerized micro-service architectures are the ideal environments to create cloud-native systems. While the service and payment/revenue model are the same in both migrated and native scenarios, the total cost of ownership (TCO) is substantially different due to the migration costs. Rationally, SPs should offer their software at a higher price to compensate for their migration costs, however this may not always be competitively feasible or desirable.

While architectural challenges in migration have been addressed [7, 33, 49, 57, 58], research exploring the link between cloud architecture and TCO, and therefore on pricing cloud services from an SP perspective is lacking. The main objective of this chapter is to extend our previous work [53] exploring the impact of two cloud architectural options, IaaS (basic virtualization) and a range of PaaS-related technology options on SPs' operating costs. We present an initial process for architecture-related cost estimation and informing pricing strategy.

This chapter is organized as follows. Section 2 reviews related work and presents the cloud migration context. Section 3 introduces the overall process. Section 4 focuses on the I/PaaS-based architecture cost calculation. In Sect. 5, we validate and illustrate our contribution using a case study. Section 6 presents different pricing structures available for SPs. The chapter concludes with a summary of contributions and suggestions for future research.

## 2 Architecture Migration Context

### 2.1 Context and Related Work

Cloud computing has attracted significant attention from the research community. Despite this, most of the research focuses on technical aspects with a limited number of studies examining the business implications of cloud adoption [36, 65]. This is somewhat surprising given the significant changes that cloud computing can generate in organizations' processes and business model, particularly for SPs [16]. Even more surprising is the lack of studies linking the value generated by cloud investments to the technical aspects of the services adopted or provided. This chapter aims to fill this gap by focusing on the impact of architectural decisions on the TCO of cloud services that SPs consume (i.e. I/PaaS) in order to provide SaaS services to their customers.

Traditionally, enterprise software was licensed under a packaged, perpetual or server license, and customers were typically required to purchase technical support and maintenance packages for a predefined period [21]. The cost of software development, production and marketing was offset against the license fees, typically paid upfront by the customer. The introduction of cloud computing accelerated the adoption of two new licensing models: subscription and utility-based licensing. The former involves an enterprise customer purchasing a license for a pre-defined time period whereas the latter involves charging the customer on a pay-per-use basis. Key advantages for the enterprise customer include (i) less upfront expenditure in licensing and (ii) no additional fees for fixes, upgrades or feature enhancements [21]. The shift from a product orientation to a service orientation is a significant disruption for SPs, not only from a strategic perspective but also from a cost- and revenue- recognition perspective, and requires in many instances a significant business model readjustment [14]. For example, cost and revenues are spread over time and producers do not receive additional fees for upgrades. Obviously, the impact of such discontinuities and shifts are not experienced by cloud-native SPs such as start-ups. Indeed, Giardino et al. [23] observe that cloud computing is particularly beneficial for start-up companies since it significantly lowers the initial investment in IT infrastructure.

It is now generally accepted that cloud computing generates a wide range of benefits and estimating the overall value generated by these type of investments is receiving growing attention from both consumers and providers [52]. Academic research has proposed a number of different approaches to estimate the business value of information technology (IT) [52]. The need for robust methodologies to assess the value generated by IT investments is driven by a trend towards value-based management, a managerial approach finalized to maximize shareholder value [5]. Value assessment techniques can be both *ex-ante* and *ex-post* [51], but it is clear that a proper *ex-ante* evaluation can better inform investment decision-making therefore potentially maximizing the return on investment or avoiding losses.

Farbey et al. [20] and Farbey and Finkelstein [19] classify value assessment methodologies in two categories:

- Quantitative/comparative methods: these typically leverage accounting methodologies to translate costs and benefits of IT investments in economic terms therefore allowing comparison between alternative investments. As such, these methods are also referred to as "objective" methods;
- Qualitative/exploratory methods: these mostly focus on the opportunities and threats that an IT investment may bring to some stakeholders. The aim in this case is to obtain an agreement over objectives through a process of exploration. These methods are also referred to as "subjective" methods given the high degree of subjectivity they may include.

Tables 1 and 2 provides a summary of different methodologies for each category as proposed by Farbey et al. [20] and Farbey and Finkelstein [19].

**Table 1.** Quantitative/Comparative methods (adapted from [51]).

| Method | Detail | Process management | Data | Features |
|---|---|---|---|---|
| Total cost of ownership (TCO) | Very detailed | Accounting and costing staff | Cost accounting and work study method | Focus on cost savings |
| Return on investment (ROI) | High | Calculation by professionals; cash flows as the aggregation of tangible cost and benefits | Cost accounting; direct and objective costs | Future uncertainty is considered; middle to high cost of implementation |
| Cost-benefit analysis | High | Carried out by experts; money values for decision makers by incorporating surrogate measures | Cost and benefit elements expressed in monetary value form | Cost-effective solutions; includes "external" and "soft" costs and benefits; numbers more important than process; high implementation cost |
| Return on management (ROM) | Low | Calculation by professionals; manipulates accounting figures to estimate the value added by management | Accounting totals (e.g. total revenue, total labor cost) | Ex-post only; no cause and effect relations can be postulated; focus on management activities; low implementation cost |
| Boundary values and spending ratios | Low | Top-down approach; senior stakeholders involved; calculation by professionals | Ratios of aggregated numbers (e.g. IT expense per employee) | Supporting benchmarking analysis; low implementation cost |
| Information economics (IE) | Very detailed | Many stakeholders involved; detailed analysis required | Ranking and rating of objectives, both tangible and intangible | All options are comprehensively dealt with; complex to implement |

For the purpose of this chapter, we focus on quantitative methods since these are the most used in practice. Among them, TCO, CBA and ROI are the most widely adopted while others like ROM, Boundary Values, Spending Ratios and Information Economics are not frequently adopted due to a perceived lower level of analysis [51] or subjectivity [63].

Despite the wide range of benefits that the adoption of cloud computing may generate for organizations, cost savings, rather than strategic return-on-investment, still represents a major factor in cloud adoption [8, 11] and TCO is *de facto* the most adopted costing model in both research and practice [52, 56]. TCO has been defined as "a procedure that provides the means for determining the total economic value of an investment, including the initial capital expenditures (*CapEx*) and the operational expenditures (*OpEx*)" [22]. The metering nature of cloud computing provides the perfect basis for extremely low-granularity TCO analysis and the opportunity to reimagine how the business value of IT is measured in both research and practice [52]. Despite its apparently simplicity and the availability of different online tools offered by cloud service providers, *ex-ante* TCO estimation is not straightforward due to the presence of long-term and hidden costs of operating in the cloud which tend to be ignored or underestimated [32]. TCO estimation frameworks used for traditional on-premise infrastructure need to be adapted to the cloud world to reflect different cost drivers [46, 62]. Rosati et al. [52] further highlight significant methodological flaws in current TCO estimation frameworks which tend to focus merely on operational cost and usually consider a small number of cost drivers.

**Table 2.** Qualitative/Exploratory methods (adapted from [51]).

| Method | Detail | Process management | Data | Features |
|---|---|---|---|---|
| Multi-Objective, Multi-Criteria (MOMC) | Any level | Top-down; consensus seeking; all stakeholders involved; best choice is computed | Priorities are stated by stakeholders; subjective evaluations of intangibles | Ex-ante; good for extracting software requirements; process is more important than numbers; selection of (a) preferred set of design goals, (b) best design alternative; high implementation cost |
| Value analysis | Any level; usually very detailed | Iterative process; senior to middle management involved; variables identified with Delphi method | Indirect; subjective evaluations of intangibles; utility scores | Ex-ante; iterative and incremental process; focus more on added value than cost saving; process is more important than numbers; high implementation cost |
| Critical success factors (CSFs) | Short list of factors | Senior management define CSFs | Interview or self-expression; quick process but requires senior management time | Ex-ante; highly selective; high implementation cost |
| Experimental methods | From detailed to abstract | Management scientists working with stakeholders | Exploratory; uncertainty reduction | Ex-ante |

From an SP perspective, this represents a major concern. Being both cloud consumers and cloud providers, properly mapping the costs of the cloud represents the basis for adequate and effective pricing strategies. SPs price their SaaS services in many ways [12]. Even though monthly or annual subscription fees is the most common pricing structure, other structures include, for example, transaction based revenue (i.e. customers are charged based on the number of transactions they perform) and premium based revenue (users are charged for premium versions besides the free versions) [13, 16, 48]. Irrespective of the pricing structure an SP adopts, a reliable estimate of the infrastructure costs it has to sustain to provide the service is required in order to ensure the existence of adequate margins [37]. This process has become more and more important for SPs due to increasing competition in the cloud environment, where SPs are sometimes forced to deliver services whose costs exceed revenues [17].

Strebel and Stage [56] applied a TCO-based decision model for business software application deployment while running simulations on hybrid cloud environments. They found that the cost-effectiveness of cloud services, from a user perspective, is positively related to the cloud-readiness of business applications and processes. The decision model they proposed was limited to a comparison of operational IT costs, such as server and storage expenses, and the external provisioning by means of cloud computing services. Li et al. [41] focused on the provider perspective. They formulated a TCO model to calculate set-up and maintenance costs (e.g. costs of hardware, software, power, cooling, staff and real-estate) of a cloud service and identified the factors involved in the utilization cost. This model consists of the total cost of all servers and resources used to provide the service. Cloud implementation and operating costs were divided into eight different categories that mainly represent fixed costs, such as set-up and maintenance costs that providers need to bear during the whole lifecycle. Han [25] presents a cost comparison between virtual managed nodes and local managed servers and storage, but neglects important cost components like licensing, training, and maintenance. Finally, Walterbusch et al. [62] presents a comprehensive TCO model for the three main cloud service models (i.e. IaaS, PaaS and SaaS), and map into their model different cost components across four phases of cloud computing i.e. initiation, evaluation, transition, operation. Costs related to system failure, backsourcing or discarding are listed but not included in the model since they are, by their nature, contingent on situation contexts and therefore difficult to translate in a mathematical formula.

Despite the large number of studies on software architecture-related factors for consideration in migration, and, likewise, the large number of studies related to TCO for cloud computing, there is a lack of papers seeking to estimate the TCO for cloud migration in conjunction with architecture concerns. The extant literature is typically focused on ex-post calculation of costs and profits independently from the wider situational context, and typically considers only cloud operational cost. For example, Andrikopoulos et al. [2] proposes a decision support system which includes a cost calculator based on per-use cost components only. Jinesh [35] presents a TCO estimation of migrating to Amazon Web Services (AWS) that includes per-use charges only. Similarly, Anwar et al. [3] examine cost-aware cloud metering for scalable services.

## 2.2 Two Migration Business Cases

Cloud computing adoption can dramatically change a company's business model and internal organization, and requires investing a significant amount of resources in the migration process. In such a context, an *ex-ante* evaluation of costs and potential benefits that such an investment may generate is crucial for effective decision-making. In this chapter, we consider two discernible business cases:

- The migration of existing legacy software and associated customers with perpetual licenses;
- Adoption of cloud-native software by new customers with no existing economic relationship with the SP.

In the first case, there is a significant post-migration discontinuity in the vendor-customer relationship and the nature of the billing. From the customer perspective, the business case can be made by comparing the as-is and the to-be solution, however this is anything but a trivial process [32]. There may be time, effort and additional hidden costs related to the migration that needs to be included in the ex-ante evaluation and recovered by both SPs and their customers [32]. In the second case, customers can make their choice on the basis of the perceived value of the service *per se*. In both cases a key consideration for SPs is the amount of cost they can sustain to generate a positive margin on their sale over a defined time period.

TCO is used to estimate the cost of cloud investments from the initial sourcing through to the end of the cloud usage, whether that is the backsourcing of information, or the client switching to other services or providers. While the measured nature of the cloud allows for a detailed *ex-post* cost analysis, *ex-ante* cost estimation can be complicated due to the uncertainty associated with multi-tenancy and resource pooling. Similarly, while there are clear cost savings in cloud computing there are also intangible cost components which are more difficult to estimate [32].

By its very nature, cloud computing enables enterprise customer scale up and down on-demand without the ties associated with a substantial upfront investment. Thus, forecasting the customer lifetime (and associated value) for a cloud customer can be difficult. Suddenly, they can leave or radically modify their usage, since switching costs in the cloud are significantly lower than on-premise. Notwithstanding this, enterprise customers and SPs require a practical approach to measuring cloud TCO.

## 3 Integrated Migration Framework and Process

Typically, a cloud migration is organized around an architectural transformation of the legacy system, independent of cost and pricing considerations. We propose an integrated process for migration planning and pricing:

Step 1: Analyze and model – Use a set of migration patterns to determine structural cloud architecture aspects;

Step 2: Right-scaling – Conduct a feasibility study to size the predicted workload to a machine (configuration) profile based on analysis of direct operational costs driven by predicted usage and experimental consumption figures;

Step 3: Right-pricing – Determine pricing for the software service based on the TCO calculation generated from the feasibility study.

## 3.1 Step 1: Analyze and Model

In the analysis and modelling step, we examine both the pre-migration context (including migration concerns) and use a set of migration patterns to determine structural cloud architecture aspects. This phase is not relevant in the context of native cloud software. For each use case, we examine the context as per Pahl et al. [49], namely:

- Setting/Application – description of the sector and classification of the application in question;
- Expectation/Drivers – the drivers and a distinction of migration benefits and expectations that potential users are aware of (their vision);
- Ignorance – factors that have been overlooked;
- Concerns – specific problems/constraints that need to be addressed.

We then conduct a multi-level analysis of requirements e.g. technology review, business analytics, migration and architecture and test and evaluation. Once this preliminary contextual analysis is completed, a set of cloud migration patterns, processes and issues as presented by Jamshidi et al. [34] and Taibi et al. [57] can be used to inform a detailed migration plan.

## 3.2 Step 2: Right-Scaling of SaaS Software

SPs seeking to migrate to the cloud need to find the right architectural configuration to meet the necessary service level agreement (SLA) obligations at the minimum cost. Therefore, a key question for a decision maker is:

*How many components can I host on a fixed cloud compute resource with a pre-defined latency performance target for a forecasted number of users of a particular application with a forecasted mix of application operation usage?*

Changes in usage require changes in the number and/or configuration of cloud resources used, which may result in additional costs. Estimation of the expected usage level or patterns is needed to predict when scaling, and related additional costs, may occur.

Furthermore, storage and networking charges are akin to commodities that can be consumed on a per-unit of usage basis. The compute costs are more difficult to predict since they are determined by the users' use of the application. In this chapter, we consider a virtual SLA-backed service that is not entirely fixed in terms of computational and storage resources allocated. Finally, the actual capacity of the offered cloud service may fluctuate over time affecting potential economies of scale and application performance. Only the cloud service provider, and not the SP, can monitor the underlying service availability thus, the first problem is right-scaling i.e., to size a predicted workload to a machine (configuration) profile. This requires usage prediction to configure IaaS or PaaS through an experimental pre-migration feasibility study, and represents the basis for an accurate estimation of operational costs. For SPs, right-scaling reduces overprovisioning and therefore usage cost of their cloud infrastructure.

## 3.3 Step 3: Right-Pricing of SaaS-Delivered Products

Monetization refers to how organizations capture value i.e. when, what and how value is converted into money [6]. Despite the fact that how SPs price and monetize their cloud offering is beyond the scope of the TCO process adopted in this chapter, it is important to understand as the TCO represents a critical component of SPs' pricing decision. A monetization framework for SPs usually comprise three models, namely:

- Architecture model: the source and target architecture need to be considered together with planned changes in functional or non-functional properties;
- Cost model: the expected direct operational costs need to be estimated including basic infrastructure and platform costs, additional features for external access and networking, internal quality management, and development and testing costs, and mapped into the TCO estimation;
- Revenue model: expected revenues based on a selected pay-per-use or subscription model.

From an SP perspective, the relationship between cloud cost and price (P) can represented as follows:

$$P = TCO \times (1 + \mu) \qquad (1)$$

Where $\mu$ represents the percentage of profit the producer aims to obtain. Understanding how SaaS usage translates in to IaaS costs is of primary importance for SPs since the SaaS income should cover the corresponding infrastructure costs. The interplay between these three models ultimately determines the attractiveness of the cloud offering of an SP in the marketplace. In this context, relevant questions to consider are:

- Which factors are static and might be considered as a baseline for the cost calculation?
- What are the additional costs for scaling up beyond the baseline?
- What is the best combination of cost and revenue model that maximize profit in the short- and long-term?

## 3.4 Total Cost of Ownership and Cost Factors

TCO, in a strict sense, is the sum of the initial investment required to purchase an asset (*CapEx*) plus the operating costs that the cloud generates (*OpEx*). When choosing among alternatives, SPs should look at both components of TCO to evaluate the investment properly. Migration costs tend to be omitted in cloud TCO estimations even though they can be substantial and change the overall return on investment. TCO calculation can be formalized as follows:

$$TCO = CapEx + OpEx \tag{2}$$

In the context of our study, $OpEx$ includes fixed (e.g. location and size) and variable (i.e. usage) IaaS cost components while $CapEx$ includes migration and implementation costs (e.g. development and testing, project management etc.). Walterbusch et al. [62] provide a comprehensive list of cost components that may be considered for estimating TCO of SP cloud migration.

In order to estimate the cost associated with the expected SaaS usage, we consider costs at the SP level. In terms of IaaS operational costs for an SP we focus on compute, storage and network resources since they usually represent the most significant cost components. IaaS costs can be categorized as either (i) fixed (size of the reserved/allocated resources, availability, location, and other supplemental and/or premium services) or (ii) variable (i.e., usage of all respective IaaS resources). Like other fixed cost factors, reconfiguration is possible, but not considered in this chapter. Availability is considered as a contractually guaranteed property and it is also assumed to be fixed.

## 4  I/PaaS Cost Calculation Process

The nature of the cloud makes it difficult to determine the input variables of the TCO model, but, as we will see, architecture quality concerns such as performance and availability can drive this process. Cloud architecture qualities, and corresponding costs, can be influenced by compute, storage and network resources. Therefore, a reliable TCO estimation requires at least two mappings from SaaS (service provided) to I/PaaS (service consumed): (i) map SaaS to I/PaaS metrics in order to link expected (SLA) and actual level of quality; and (ii) map SaaS to I/PaaS usage patterns in order to link SaaS usage variation to the required level of I/PaaS resources. Figure 1 summarizes the cost estimation process that we will now apply.

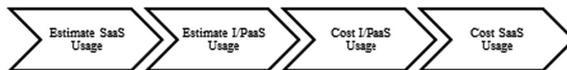

**Fig. 1.** Costing SaaS usage - estimation process [53].

### 4.1  Cost Estimation Process

In a cloud migration scenario, an SP needs to migrate the system architecture of the target on-premise software product and change the corresponding cost and revenue models at the same time. As highlighted before, the new models heavily depend on expected or predicted usage, both of which are difficult to estimate. In fact, any estimation of SaaS usage volumes will determine IaaS usage requirements but customers' usage can be subject to temporary peaks that might generate spikes in costs due to ineffective IaaS usage.

Estimation complexity varies between the two business cases identified earlier, i.e. migrated or cloud-native application. Usage patterns of the existing customer base can be determined with reasonably high accuracy, as opposed to the future behavior of an unknown customer cohort in the cloud-native scenario. The initial two phases relate to usage estimation at both the SaaS and IaaS level. SaaS usage can be mapped onto IaaS by experimental means using feasibility studies or other mechanisms. A third phase is concerned with IaaS cost estimation, which is driven by the usage estimation and SLA obligations. IaaS configuration heuristics can be used to identify the most efficient infrastructure configuration. The fourth and final phase is related to pricing the SaaS service based on the outcome of the previous stages.

## 4.2 Architecture Selection and Cost/Revenue Prediction

From an SP perspective, the list of selection criteria of a cloud provider includes both fees and the associated billing model. Many IaaS providers offer monthly basic subscription fees with additional fees for premium services such as scalability, access monitoring (e.g., IP endpoint, network bandwidth), and advanced self-management. An SP requires a clear comparison of costs and revenues resulting from the cloud adoption. This has to be an "apples to apples" comparison [32]. Even though we primarily discuss IaaS, similar assumptions can be made for PaaS services. PaaS-level costs need to address both development and deployment and need to be aligned with SaaS-level income. In order to determine a profitable and sustainable pricing model, the following steps need to be taken:

- Estimation of the TCO of consumed cloud services on the basis of the expected usage of the provided SaaS service;
- Estimation of the expected level of revenues on the basis of expected usage of suitable fees level;
- A sensitivity analysis of I/PaaS costs to potential changes in SaaS usage;
- Assessment of the alignment of the selected pricing model with the market strategy of the SP;
- Assessment of the sustainability of the selected pricing model both in the short- and long-term.

## 4.3 Assumptions – Resource Cost Modeling and Right-Scaling

In order to make this more practically relevant, we can look at the different resource types and compare them in terms of utilization and cost fluctuations in common deployments (and resulting impact on cost estimation). Cost modeling for compute versus storage services are fundamentally different. Storage usage is more predictable and current cloud service pricing models support a commodity-style costing. Compute usage and related cost is more complicated to predict since it can fluctuate significantly over time and contributes disproportionately to the achievement of economies of scale.

SPs need to make configuration assumptions which may or may not prove to be accurate. Scenario analysis may help to achieve better estimation.

For illustration purposes, a simple initial configuration of IaaS resources could be based on 80% reserved and 20% on-demand instances. This combines reliable core provisioning without overprovisioning for extra demand (in which case on-demand instances are acquired). The benefits of this strategy are:

- 60–80% utilization of used instances is achievable if the reserved instances deal with peak demand;
- Up to 50% cost reduction compared to on-demand instances only.

Another factor impacting resource requirement is the nature of the architecture. Stateless, loosely-coupled architectures help accommodate extra demand and enable scalability by just using additional resources on-demand without much start-up costs (transfer of state to other resources).

### 4.4   An Exemplar Costing Model

In order to understand pricing models of IaaS and PaaS providers, we report exemplar categories and common pricing models (Table 3). This is largely built on Microsoft Azure pricing information, but is typical of other providers. Relevant costing models focus primarily on storage in GB and transactions (read/write). A proper estimation of IaaS costs associated with a SaaS application provisioning is needed in order to (i) select the technically best option, and (ii) estimate the costs for hosting the SaaS application, for example, in a PaaS cloud. Quality concerns other than the expected workload (e.g. availability expectations, failover strategy etc.) have to be considered in the process as well. Effectively, the estimation process needs to include the number of storage units and total size as an input, and the costs, estimated over a defined period, with predicted growth, and for different replication options as an output.

A further complication is that pricing models between platform providers are difficult to compare due to different definitions of price components. Consequently, a formal and clear estimation framework for an economic evaluation of different solutions to deliver a SaaS service is needed.

## 5   Illustration and Validation – Case Study

We now illustrate the estimation process presented in Sect. 4 using a case study. The estimation process was applied to an SP migrating a legacy client-server on-premise single-tenant enterprise application to the cloud by re-designing, re-engineering and recoding the system as a cloud application. The SP is a small-medium enterprise which provides a document management application. Its application has over 1,000 existing client installs and in this case study, we present the TCO estimation of migrating 240 of these to the new cloud platform over a 3-year period. The main business requirements for the SP to adopt the cloud were (i) to pursue flexibility across different devices and situational contexts, and (ii) to increase the customer base through efficient entry in to new geographical markets. The solution requires meeting high-volume data storage and processing needs.

**Table 3.** Storage cost component (adapted from [53]).

| Component | Description |
|---|---|
| Region | A region is a set of datacenters deployed within a latency-defined perimeter and connected through a dedicated regional low-latency network |
| Replication | Cloud providers usually create multiple copies of each database in order to ensure durability and high availability. Cloud users can choose the replication option that best fits its needs but each option come with different a different price. Sample configurations include:<br>• Local Redundant – a number of copies are stored in the same data-center and region of the storage account, but across different fault or upgrade domains<br>• Zone Redundant – a number of copies are stored in different data-centers, which have slightly less throughput than Local redundancy<br>• Geo Redundant – a number of copies are stored in different data-centers, with a back-up, separate multiple saves in a specific secondary region to allow to recover from potential region failure<br>• Read-Only Geo Redundant – Similar to geo redundancy with read access to secondary data<br>All replication operations are done asynchronously |
| Size | Storage cost is positively related with the volume of data stored in a database |
| Transactions | Storage cost depends on the number of transactions - i.e. read/write blob operations – performed in each database. The higher the number of transactions, the higher the cost |
| Data transfer | Storage cost is positively related with volume of data being transferred from/to the database. However, the cost of data transfer is usually charged only when data is moved out from the geographical region where it was stored. In-region transfers are usually free |

## 5.1 Application Overview

The case site is a small-to-medium sized SP that overs document management services to the logistics sector. The application is a Document Management System (DMS), which enables a user to scan paper documents from enterprise-grade scanners and save them on a cloud store as electronic images. Documents are classified under custom types, such as invoice or delivery docket, and specific metadata templates are used to store search-able tagged data against the documents for future retrieval and reporting. The SP wishes to deploy the software in the cloud and due to the commercially sensitive nature of the documents being scanned, data location is major concern. The SP does not have enough information on the cost of migration and cloud deployment specifically to inform a migration decision and/or pricing strategy. Specifically:

- Technology review - the SP has network concerns regarding the upload and download data transfer speeds and services for in-cloud document processing.

- Business analysis – the SP has concerns about security and data privacy regulations e.g. GDPR.
- Migration and architecture – the preferred solution is a two-stage incremental migration plan (IaaS and PaaS) to migrate document scanning, storage and processing to a scalable cloud architecture.
- Test and evaluation – scalability, performance, integration and security must meet agreed criteria.

A summary migration plan with stepwise migration from on-premise via IaaS into a PaaS cloud could be implemented as follows:

1. IaaS Compute Architecture: The application can be packaged in-to VMs. License fees for components of the application are incurred as usual. The business problem is scaling out; adding more VMs means adding more license fees for every replicated component. From a technical point of view, multiple copies of data storage that are not in sync might cause integrity problems.
2. DaaS Storage: Refactor and extract storage i.e. use a virtual data-as-a-service (DaaS) solution for storage needs. This alleviates the technical integrity problem cited above.
3. PaaS Cloud Data Storage: Package the whole DBMS into a single virtual machine. This alleviates the business license fee problem for the DBMS and simplifies data management, but other license fees may still occur.
4. Full Application Migration: Migrate to a PaaS service. Apart from solving technical problems, this significantly mitigates the licensing fees issue.

Ultimately and for the purposes of this case, the application has been redesigned and coded specifically to run as a cloud application on the Microsoft Azure public cloud platform.

## 5.2 TCO Calculation

The TCO is made up of the implementation costs of the new cloud application and the cloud charges incurred in running the new system on Microsoft Azure. Estimated implementation costs (*CapEx*) were classified into seven implementation phases: Business Analysis, Cloud Architecture Design, Data Design, Security Framework Design, Development and Test (see Table 10), Performance and Costs Analysis (see Tables 11, 12 and 13). It should be noted that the calculations do not include the operational costs of migrating the customers to the new cloud web application.

The application is a multi-process system since it comprises a web server compute resource and a separate image processing compute resource. However, the functional dependency between these do not need to be considered in the TCO analysis since the image processing worker VM acts completely asynchronously to the web server role web requests which continue regardless of the state of the image processor. Therefore, we have calculated the multi-tenant VM requirements based on a simple linear multiplication of the CPU load per tenant.

IaaS usage charges (*OpEx*) are estimated considering the two most relevant cost components:

- A cloud data store – made up of a NoSQL Table structure (using the Microsoft Azure Table service) and an object store (using the Microsoft Azure Blob Storage service). Table and blob storage are platform services that allow a more fine-grained costing. As such, these need to be considered on an individual service base.
- A cloud compute architecture – made up of a separate compute resource for the web server of the web application (Web Role Virtual Machine), and a separate compute component for carrying out the image processing functions, such as barcode reading (Worker Role Virtual Machine).

Our calculation is based on the Microsoft Azure services pricing reported in Tables 4, 5, and 6. In order to forecast the usage of cloud storage resources, we used actual historical data over an eleven-month period from an existing average-sized tenant with a typical application usage pattern. To estimate the computing resources required, we monitored the usage and performance statistics during a snapshot of the operational use of the application by the same typical user. Tables 7, 8, and 9 summarize the usage profile adopted in the calculation.

**Table 4.** Blob storage prices (adapted from [53]).

| Service | Redundancy | Cool tier price | General purpose price |
|---|---|---|---|
| Price per GB/Month space | Local | € 0.013 | € 0.020 |
| | Geo | € 0.025 | € 0.041 |
| Price per 10,000 transactions | Local | € 0.084 | € 0.003 |
| | Geo | € 0.169 | € 0.003 |
| Price per GB data access write | Local | € 0.002 | - |
| | Geo | € 0.004 | - |

**Table 5.** Table storage prices (adapted from [53]).

| | Redundancy | Price |
|---|---|---|
| Price per Entity/GB/Month | Local redundant | € 0.059 |
| | Geo redundant | € 0.085 |
| Price per 10,000 transactions (PUT) | Local redundant | € 0.003 |
| | Geo redundant | € 0.003 |

**Table 6.** Compute prices (adapted from [53]).

| VM type | No. of CPU cores | Annual cost Azure VM (€) | VM type | No. of CPU cores | Annual cost Azure VM (€) |
|---|---|---|---|---|---|
| a1 | 1 | 598.18 | d4 | 8 | 8,936.93 |
| a2 | 2 | 1,205.28 | d1 v2 | 1 | 1,107.07 |
| a3 | 4 | 2,401.63 | d2 v2 | 2 | 2,232.00 |
| a4 | 8 | 4,812.19 | d3 v2 | 4 | 4,464.00 |
| d1 | 1 | 1,107.07 | d4 v2 | 8 | 8,936.93 |
| d2 | 2 | 2,232.00 | d5 v2 | 16 | 17,873.86 |
| d3 | 4 | 4,464.00 | d2 v3 | 2 | 1,589.18 |

**Table 7.** Usage profile of a typical tenant (adapted from [53]).

| Items | Size |
|---|---|
| Total number of scanned documents per annum | 145,853 |
| Average number of document table entities per month | 14,675 |
| Number of peak entities per day | 3,551 |
| Number of peak entities per hour | 1,137 |
| Average table entity size (in bytes) | 2,160 |
| Average scanned image file size (in Kilobytes) | 666 |
| Average template file size (in bytes) | 2,200 |

**Table 8.** Forecasted input parameters (adapted from [53]).

| Per tenant | End of year | | |
|---|---|---|---|
| | 1 | 2 | 3 |
| Number of documents | 176,105 | 352,210 | 528,314 |
| Document table size (in Gigabytes) | 0.380 | 0.761 | 1.141 |
| Number of image blobs | 176,105 | 352,210 | 528,314 |
| Image blobs size (in Gigabytes) | 117 | 235 | 352 |
| Document template file blobs | 2 | 3 | 6 |
| Total template blob storage (in bytes) | 4,400 | 8,800 | 13,200 |

**Table 9.** Summary parameter values (adapted from [53]).

| Workload | % |
|---|---|
| Web role peak CPU load | 67.1% |
| Web role average CPU load | 31.5% |
| Worker role peak CPU load | 24.3% |
| Worker role average CPU load | 10.4% |

**Table 10.** Migration and implementation costs (adapted from [53]).

| Implementation phase | Cost (€) |
|---|---|
| Implementation consultancy costs – business analysis (Contract hours) | 16,078 |
| Implementation consultancy costs – security design (Contract hours) | 27,237 |
| Implementation consultancy costs – design and development (Contract hours) | 80,662 |
| Project management and implementation design (Staff Salaries) | 16,265 |
| Development and Testing (Staff Salaries) | 17,465 |
| Non-staff or non-contractor costs (Cloud Testbed subscription, test equipment, travel) | 10,940 |
| Total | 168,647 |

## 5.3 Experimentation – Usage and Cost

Table 10 summarizes the estimated implementation and migration costs for the SP (€168,647). The most significant cost component, which represents 47.83% of the overall migration costs, is by far consultancy costs for design and development, followed by security design (16.15%). Such a significant amount of upfront migration costs further highlights the need to include such costs into TCO estimation to inform both adoption and pricing decisions.

Tables 11, 12, and 13 summarize IaaS usage costs estimated as a linear combination of usage parameters and price of each service. Note that these pragmatic/empirical observations stem from experiments in a live feasibility study and have been implemented on the basis of the following assumptions:

**Table 11.** Blob storage costs (adapted from [53]).

| Costs per tenant | Space cost (€) | | Transactions cost (€) | |
|---|---|---|---|---|
| Redundancy | Local | Geo | Local | Geo |
| End year 1 | 8.87 | 17.80 | 1.48 | 2.97 |
| End year 2 | 26.60 | 53.41 | 1.48 | 2.97 |
| End year 3 | 44.33 | 89.02 | 1.48 | 2.97 |
| | Data access write cost (€) | | Total cost (€) | |
| Redundancy | Local | Geo | Local | Geo |
| End year 1 | 1.48 | 2.96 | 11.83 | 23.73 |
| End year 2 | 4.43 | 8.87 | 32.52 | 65.25 |
| End year 3 | 7.39 | 14.78 | 53.21 | 106.77 |

*Note: Blob storage costs for template files were ignored due to their negligible amount.*

**Table 12.** Table storage costs (adapted from [53]).

| Costs per tenant | Space Cost (€) | | Transactions Cost (€) | | Total Cost (€) | |
|---|---|---|---|---|---|---|
| Redund. | LR | GR | LR | GR | LR | GR |
| End year 1 | 0.13 | 0.19 | 0.05 | 0.05 | 0.19 | 0.25 |
| End year 2 | 0.40 | 0.58 | 0.05 | 0.05 | 0.46 | 0.63 |
| End year 3 | 0.67 | 0.97 | 0.05 | 0.05 | 0.73 | 1.02 |

*Note: LR (Local Redundant); GR (Geo Redundant); Redund. (Redundancy)*

**Table 13.** Compute costs (adapted from [53]).

| End year | Clients migrated | Number of VMs (WeR) | Number of VMs (WoR) | Storage costs (LR) (€) |
|---|---|---|---|---|
| 1 | 80 | 6 | 2 | 946 |
| 2 | 80 | 18 | 4 | 3,548 |
| 3 | 80 | 30 | 6 | 7,805 |
| | | Storage costs (GR) (€) | Compute costs (WS) (€) | Compute costs (IP) (€) |
| 1 | 80 | 1,898 | 9,536 | 3,179 |
| 2 | 80 | 7,118 | 28,606 | 6,357 |
| 3 | 80 | 15,660 | 47,676 | 9,536 |

*Note: WeR (Web Role); WoR (Worker Role); LR (Local Redundant); GR (Geo Redundant); WS (Web Server VMs); IP (Image Processing VMs).*

- The existing deployment does not include any data caching which would obviously reduce the CPU overhead and data storage access costs.
- There is no optimization of the queries to the table service to optimize CPU load over the TCO estimation period.
- There is no performance tuning on the application and/or on the platform during the TCO estimation period.
- There is no smoothing effect of multiple tenants sharing the same application compute resources.

The use case we present in this chapter involves a significant image-processing component resulting in high upload- and download- volumes and the in-cloud processing of images. The most critical challenge at the architectural level was to select the optimal Virtual Machine type from the available types on the Microsoft Azure platform; we carried out a benchmark study of the performance of the different "flavors" of the role VMs when running the data layer functions of the new application. The costs presented in Tables 11, 12, and 13 are based on the D2-V3 VM type which represented

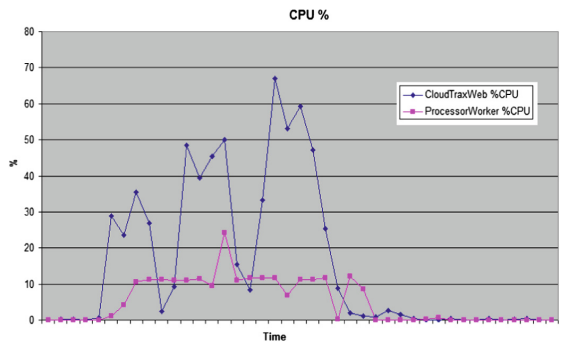

**Fig. 2.** Compute usage over a twenty-minute monitoring period [53].

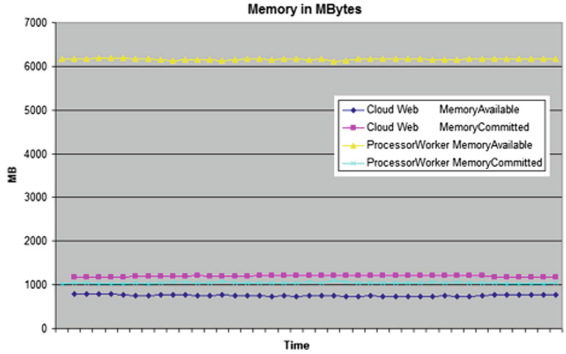

**Fig. 3.** Storage usage over a twenty-minute monitoring period [53].

the best trade-off between TCO and SLA requirements on the basis of the average tenant usage.

Among different TCO components, compute is by far the most significant (€129,701), and also the most fluctuating resource (see Fig. 2). As such, its efficient and effective usage should be the main concern of the SP. Storage, as predicted, is relatively stable and predictable with essentially fixed costs (see Fig. 3), and accounts for a very tiny portion of the TCO (€293.31 – 0.001%). The heavy image processing results in higher-than-normal network bandwidth and storage requirements. As a consequence, the observations should also hold for applications with less data volume and would thus cover the majority of typical transactional business applications.

## 6 Right-Pricing of SaaS Service

Once a SP has established the costs of cloud delivery including compute, storage, and migration, if appropriate, the price can be determined using Eq. 1 as outlined in Sect. 3.3.

At this point in time, the SP typically must decide on their pricing strategy driven by their overall strategic objectives i.e. determine the value for μ. The selection of an appropriate pricing strategy is increasing seen as a source of competitive advantage thus right-pricing is crucial for the SP, in the cloud or otherwise [30].

There are a number of pricing strategies that the SP can choose from, the most common strategies being variants or combinations of cost-based, demand-driven or value-based, and competition-oriented [29, 38]. Cost-based strategies determine the price level using cost accounting. Harmon et al. [26] suggest that these approaches are short-term, tactical in nature, and place the interests of the seller over the interests of the buyer leading to overpricing in weak markets and underpricing in strong markets. In contrast, demand-driven or value-based costing recognizes the price that a customer is willing to pay, mostly, depends on the customer's value requirements, not the SP. For Harmon et al. [26], the goal of value-based pricing is to enable more profitable pricing by capturing more value which in turn should input, if not determine, the level of product (development) costs that the company is willing to incur or not. While commentators suggest that this is the best overall approach to take [29, 31], it is not without drawbacks. Hinterhuber [29] notes the difficulty in obtaining and interpreting the necessary data to measure customer value and that in some cases, value-based pricing can lead to relatively high prices. Competition-oriented pricing is based on anticipated or observed price levels of competitors for determining price points [29]. The weakness in competition-based pricing is that again customer willingness to pay or costs are not necessarily taken in to account [29]. Each of these pricing strategies are prevalent in cloud computing [1]. It should be noted that profitability may or not be a goal in initial pricing strategies. SPs may offer unprofitable software services (including zero pricing) for a variety of reasons in order to drive market expansion or maintain customer satisfaction levels [30, 59]. As such, $\mu$ may be negative. For each pricing strategy outlined, TCO remains a useful calculation and indeed can help address the drawbacks in each strategy.

For our purposes, right pricing is combinatorial approach taking in to account the costs of cloud deployment but also scalability. Scalability, in this context, represents future customer demand. $\mu$ therefore becomes a variable that can be used to support the testing all pricing strategies at different levels through scenario analysis or even lean startup methodologies. Additionally, once a pricing strategy has been decided, a specific pricing structure must be agreed e.g. pay-per-use, annual or monthly subscription per user etc.

## 7   Conclusions and Future Developments

Our literature review highlighted a clear lack of processes integrating software architecture and costing within a cloud migration scenario. This chapter aims to fill such a gap by investigating the link between architectural decisions and the impact on costing in cloud migration and therefore making an initial contribution in this context [42]. Specifically, we have identified the major determinants of SaaS usage costs and integrated them into one single process to estimate the corresponding I/PaaS costs. This would represent the basis for defining the pricing a SaaS licensing model, and ultimately impact the profit margins of an SP. Due to the differences in factors and account types between the IaaS/PaaS providers, a generic, formalized model cannot exist. Thus, our aim was to identify the factors influencing this calculation and to illustrate this through a real-life case study.

As no single formula to easily determine right-scaling and right-pricing was identified in our literature review, in this chapter we propose an initial process for estimating operating costs and dependencies, and architecture-related costs.

Cloud adoption, like all technology investments, results in direct tangible costs such as cloud resources but also in intangible costs, e.g., change management, vendor management, risk mitigation etc. [47]. In our case study, we have moved beyond merely operating costs by including some of these indirect cost components. However, our example does not aim to provide a comprehensive list of such costs. Furthermore, the research presented in this chapter is subject to a series of limitations which curtail its generalizability, but it also presents avenues for future research. First, we have focused on a business-to-business SP targeting small and medium enterprises, and to a single cloud service provider. As such, our conclusion is not directly generalizable to business-to-consumer SPs. Further studies may account for more complex models suitable for larger and more mature organizations or may seek to compare functionality, quality and costs across multiple providers [24].

Second, we did not consider recent developments in cloud architectures like container technology and microservices architectures, which are an increasing feature in the enterprise cloud and enabling new provisioning and payment models, new services like serverless computing (also referred to as 'function-as-a-service'), which will radically change how SPs conceptualize costs and pricing. The adoption of serverless computing, for example, is growing significantly in order to increase efficiencies and provisioning speeds. This relatively new paradigm of cloud computing envisages a model of computing where effectively all resources are pooled including hardware, operating systems and runtime environments [28]. As a result, an SP only concerns themselves with relatively lightweight, single purpose stateless functions that can be executed on demand without consuming any resources until the point of execution. The serverless paradigm introduces greater separation of concerns between cloud service providers and SPs to the extent that much more responsibility is transferred to the cloud service provider. In addition, the SP benefits from much less complexity but also benefits from a lower cost of deployment related to the lightweight nature of functions and by cloud service pricing driven at the level of execution runtime for computer code rather than how long an instance is running [18]. The market for serverless computing is expected to grow to US\$7.72 billion by 2021 [44]; as such, it is not surprising that many of the major cloud service providers have entered the market including AWS (Lambda), Microsoft (Azure Function), Google (Cloud Function), and IBM (Bluemix OpenWhisk). Research on serverless computing is at a very early stage of development and is primarily based on AWS Lambda [43]. While most of the research is focused on use cases, Lynn et al. [43] report a small number of studies that report cost efficiencies resulting from serverless implementations [40, 60, 61]. Given the novelty of serverless computing, the novelty of serverless pricing models, emerging use cases, and the dearth of research on business value and serverless migration, this area would seem to be a fruitful area for research moving forward. As other novel cloud services emerge, there will be a need for business value research, and TCO research specifically, not least fog computing [10], edge computing [54], cloud service brokerage and enterprise app marketplaces [50], quantum computing as a service [55], and self-organizing self-managing heterogeneous clouds [64].

Our work shows that there is a need for an integrated perspective accommodating architecture and cost in order to provide a clear basis for service pricing and revenue, and that the traditional TCO approaches cannot be applied without adaptation. Even though this chapter focuses on TCO, the same need for adaptation applies to other value assessment methodologies. As such, they present additional avenues for future research. Our chapter also highlights the need for collaboration between business, accounting and computer science researchers. As businesses become more and more reliant on cloud computing, such a collaboration is essential for providing a comprehensive understanding of the financial implications of adopting specific software architectures in the cloud computing context. This likely requires not only adaptation in common activity-based and resource-based costing methodologies but also in software and systems design.

**Acknowledgements.** The research work described in this chapter was supported by the Irish Centre for Cloud Computing and Commerce, an Irish National Technology Centre funded by Enterprise Ireland and the Irish Industrial Development Authority.

# References

1. Al-Roomi, M., Al-Ebrahim, S., Buqrais, S., Ahmad, I.: Cloud computing pricing models: a survey. Int. J. Grid Distrib. Comput. **6**(5), 93–106 (2013)
2. Andrikopoulos, V., Song, Z., Leymann, F.: Supporting the migration of applications to the cloud through a decision support system. In: IEEE Sixth International Conference on Cloud Computing (2013)
3. Anwar, A., Sailer, A., Kochut, A., Schulz, C.O., Segal, A., Butt, A.R.: Cost-aware cloud metering with scalable service management infrastructure. In: IEEE 8th International Conference on Cloud Computing, pp. 285–292 (2015)
4. Armbrust, M., et al.: A view of cloud computing. Commun. ACM **53**(4), 50–58 (2010)
5. Arnold, G., Davies, M.: Value-Based Management: Context and Application. Wiley, New York (2000)
6. Baden-Fuller, C., Haefliger, S.: Business models and techno-logical innovation. Long Range Plan. **46**(6), 419–426 (2013)
7. Balalaie, A., Heydarnoori, A., Jamshidi, P., Tamburri, D.A., Lynn, T.: Microservices migration patterns. Softw. Pract. Exp. **48**, 1–24 (2018)
8. Bain and Company: The Changing Faces of the Cloud (2017). http://www.bain.com/publications/articles/the-changing-faces-of-the-cloud.aspx. Accessed 28 Jan 2018
9. Berman, S.J., Kesterson-Townes, L., Marshall, A., Srivathsa, R.: How cloud computing enables process and business model innovation. Strategy Leadersh. **40**(4), 27–35 (2012)
10. Bonomi, F., Milito, R., Zhu, J., Addepalli, S.: Fog computing and its role in the internet of things. In: Proceedings of the First Edition of the MCC Workshop on Mobile Cloud Computing, pp. 13–16 (2012)
11. CFO Research: The Business Value of Cloud Computing: A Survey of Senior Finance Executives. CFO Publishing (2012). http://lp.google-mkto.com/rs/google/images/CFO%2520Research-Google_research%2520report_061512.pdf. Accessed 20 Jan 2018
12. Cusumano, M.A.: The changing labyrinth of software pricing. Commun. ACM **50**(7), 19–22 (2007)
13. Cusumano, M.A.: The changing software business: moving from products to services. Computer **41**(1), 20–27 (2008)
14. DaSilva, C.M., Trkman, P., Desouza, K., Lindic, J.: Disruptive technologies: a business model perspective on cloud computing. Technol. Anal. Strateg. Manag. **25**(10), 1161–1173 (2013)
15. Dillon, T., Wu, C., Chang, E.: Cloud computing: issues and challenges. In: IEEE International Conference on Advanced Information Networking and Applications, pp. 27–33 (2010)
16. D'souza A., Kabbedijk, J., Seo, D., Jansen, S., Brinkkemper, S.: Software-as-a-service: implications for business and technology in product software companies. In: Pacific Asia Conference on Information Systems (2012)
17. Durkee, D.: Why cloud computing will never be free. Commun. ACM **53**(5), 62–69 (2010)

18. Eivy, A.: Be wary of the economics of "serverless" cloud computing. IEEE Cloud Comput. **4**(2), 6–12 (2017)

19. Farbey, B., Finkelstein, A.: Evaluation in software engineering: ROI, but more than ROI. Working Paper Series - Department of Computer Science University College London – LSE, (2000). http://is.lse.ac.uk/all_wp.htmS

20. Farbey, B., Land, F., Targett, D.: How to Assess Your IT Investment: A study of Methods and Practice. Butterworth-Heinemann, Oxford (1993)

21. Ferrante, D.: Software licensing models: what's out there? IT Prof. **8**(6), 24–29 (2006)

22. Filiopoulou, E., Mitropoulo, P., Tsadimas, A.: Integrating cost analysis in the cloud: a SoS approach. In: 11th International Conference on Innovations in Information Technology (IIT) (2015)

23. Giardino, C., Bajwa, S.S., Wang, X., Abrahamsson, P.: Key challenges in early-stage software startups. In: Lassenius, C., Dingsøyr, T., Paasivaara, M. (eds.) XP 2015. LNBIP, vol. 212, pp. 52–63. Springer, Cham (2015). https://doi.org/10.1007/978-3-319-18612-2_5

24. Gilia, P., Sood, S.: Automatic selection and ranking of cloud providers using service level agreements. Int. J. Comput. Appl. **72**(11), 45–52 (2013)

25. Han, Y.: Cloud computing: case studies and total costs of ownership. Inf. Technol. Libr. **30**(4), 198–206 (2011)

26. Harmon, R., Demirkan, H., Hefley, B., Auseklis, N.: Pricing strategies for information technology services: a value-based approach. In: Hawaii International Conference on System Sciences (HICSS), pp. 1–10 (2009)

27. Heilig, L., Voß, S.: Decision analytics for cloud computing: a classification and literature review. In: Bridging Data and Decisions, pp. 1–26 (2014)

28. Hendrickson, S., Sturdevant, S., Harter, T., Venkataramani, V., Arpaci-Dusseau, A.C., Arpaci-Dusseau, R.H.: Serverless computation with openlambda. Elastic **60**, 1–7 (2016)

29. Hinterhuber, A.: Customer value-based pricing strategies: why companies resist. J. Bus. Strategy **29**(4), 41–50 (2008)

30. Hinterhuber, A., Liozu, S.M.: Is innovation in pricing your next source of competitive advantage? Bus. Horiz. **57**, 413–423 (2014)

31. Ingenbleek, P., Debruyne, M., Frambach, R.T., Verhallen, T.M.: Successful new product pricing practices: a contingency approach. Mark. Lett. **14**(4), 289–305 (2003)

32. ISACA: Calculating Cloud ROI: From the Customer Perspective (2012). https://www.isaca.org/knowledge-center/research/researchdeliverables/pages/calculating-cloud-roi-from-the-customer-perspective.aspx. Accessed 20 Jan 2018

33. Jamshidi, P., Ahmad, A., Pahl, C.: Cloud migration research: a systematic review. IEEE Trans. Cloud Comput. **1**(2), 142–157 (2013)

34. Jamshidi, P., Pahl, C., Chinenyeze, S., Liu, X.: Cloud migration patterns: a multi-cloud service architecture perspective. In: International Workshop on Engineering Service Oriented Applications – WESOA 2014 (2014)

35. Jinesh, V.: Migrating your existing applications to the AWS cloud. A Phase-driven Approach to Cloud Migration (2010). http://docs.huihoo.com/amazon/aws/whitepapers/Migrating-your-Existing-Applications-to-the-AWS-Cloud-October-2010.pdf. Accessed 21 Jan 2018

36. Karunakaran, S., Krishnaswamy, V., Rangaraja, P.S.: Business view of cloud: decisions, models and opportunities–a classification and review of research. Manag. Res. Rev. **38**(6), 582–604 (2015)

37. Laatikainen, G., Ojala, A.: SaaS architecture and pricing models. In: IEEE International Conference on Services Computing (SCC), pp. 597–604 (2014)

38. Lehmann, S., Buxmann, P.: Pricing strategies of software vendors. Bus. Inf. Syst. Eng. **1**(6), 452–462 (2009)

39. Leimbach, T., et al.: Potential and Impacts of Cloud Computing Services and Social Network Websites. Science and Technology Options Assessment (STOA) (2016). http://www.europarl.europa.eu/RegData/etudes/etudes/join/2014/513546/IPOL-JOIN_ET(2014)513546_EN.pdf. Accessed 15 Aug 2016

40. Leitner, P., Cito, J., Stöckli, E.: Modelling and managing deployment costs of microservice-based cloud applications. In: Proceedings of the 9th International Conference on Utility and Cloud Computing, pp. 165–174 (2016)

41. Li, X., Li, Y., Liu, T., Qiu, J., Wang, F.: The method and tool of cost analysis for cloud computing. In: IEEE International Conference on Cloud Computing, pp. 93–100 (2009)

42. Li, H., Zhong, L., Liu, J., Li, B., Xu, K.: Cost-effective partial migration of VoD services to content clouds. In: IEEE International Conference on Cloud Computing, pp. 203–210 (2011)

43. Lynn, T., Rosati, P., Lejeune, A., Emeakaroha, V.: A preliminary review of enterprise serverless cloud computing (function-as-a-service) platforms. In: IEEE International Conference on Cloud Computing Technology and Science (CloudCom), pp. 162–169 (2017)

44. Market and Markets: Function-as-a-Service Market by User Type (Developer-Centric and Operator-Centric), Application (Web & Mobile Based, Research & Academic), Service Type, Deployment Model, Organization Size, Industry Vertical, and Region - Global Forecast to 2021 (2017). https://www.marketsandmarkets.com/Market-Reports/function-as-a-service-market-127202409.html. Accessed 2 Aug 2018

45. Marston, S., Li, Z., Bandyopadhyay, S., Zhang, J., Ghalsasi, A.: Cloud computing - the business perspective. Decis. Support Syst. **51**(1), 176–189 (1999)

46. Martens, B., Walterbusch, M., Teuteberg, F.: Costing of cloud computing services: a total cost of ownership approach. In: 45th Hawaii International Conference on System Science (HICSS), pp. 1563–1572 (2012)

47. Misra, S.C., Mondal, A.: Identification of a company's suitability for the adoption of cloud computing and modelling its corresponding return on investment. Math. Comput. Model. **53**(3), 504–521 (2011)

48. Ojala, A.: Software renting in the era of cloud computing. In: IEEE 5th International Conference on Cloud Computing (CLOUD), pp. 662–669 (2012)

49. Pahl, C., Xiong, H., Walshe, R.: A comparison of on-premise to cloud migration approaches. In: Lau, K.-K., Lamersdorf, W., Pimentel, E. (eds.) ESOCC 2013. LNCS, vol. 8135, pp. 212–226. Springer, Heidelberg (2013). https://doi.org/10.1007/978-3-642-40651-5_18

50. Paulsson, V., Morrison, J., Emeakaroha, V., Lynn, T.: Cloud service brokerage: a systematic literature review using a software development lifecycle. In: 22nd Americas Conference on Information Systems (AMCIS) (2016)

51. Ronchi, S., Brun, A., Golini, R., Fan, X.: What is the value of an IT e-procurement system? J. Purchasing Supply Manag. **16**(2), 131–140 (2010)

52. Rosati, P., Fox, G., Kenny, D., Lynn, T.: Quantifying the financial value of cloud investments: a systematic literature review. In: IEEE International Conference on Cloud Computing Technology and Science (CloudCom), pp. 194–201 (2017)

53. Rosati, P., Fowley, F., Pahl, C., Taibi, D., Lynn, T.: Making the cloud work for software producers: linking architecture, operating cost and revenue. In: 8th International Conference on Cloud Computing and Services Science (CLOSER) (2018)

54. Shi, W., Cao, J., Zhang, Q., Li, Y., Xu, L.: Edge computing: vision and challenges. IEEE Internet Things J. **3**(5), 637–646 (2016)

55. Singh, H., Sachdev, A.: The quantum way of cloud computing. In: International Conference on Optimization, Reliability, and Information Technology (ICROIT), pp. 397–400 (2014)

56. Strebel, J., Stage, A.: An economic decision model for business software application deployment on hybrid cloud environments. Multikonferenz Wirtschaftsinformatik, MKWI (2010)

57. Taibi, D., Lenarduzzi, V., Pahl, C.: Processes, motivations and issues for migrating to microservices architectures: an empirical investigation. IEEE Cloud IEEE Cloud Comput. J. **4**(5), 22–32 (2017)
58. Taibi, D., Lenarduzzi, V., Pahl, C.: Architectural patterns for microservices: a systematic mapping study. In: 8th International Conference on Cloud Computing and Services Science (CLOSER) (2018)
59. Terho, H., Suonsyrjä, S., Karisalo, A., Mikkonen, T.: Ways to cross the rubicon: pivoting in software startups. In: Abrahamsson, P., Corral, L., Oivo, M., Russo, B. (eds.) PROFES 2015. LNCS, vol. 9459, pp. 555–568. Springer, Cham (2015). https://doi.org/10.1007/978-3-319-26844-6_41
60. Villamizar, M., et al.: Cost comparison of running web applications in the cloud using monolithic, microservice, and AWS lambda architectures. SOCA **11**(2), 233–247 (2017)
61. Wagner, B., Sood, A.: Economics of resilient cloud services. In: IEEE International Conference on Software Quality, Reliability and Security Companion (QRS-C), pp. 368–374 (2016)
62. Walterbusch, M., Martens, B., Teuteberg, F.: Evaluating cloud computing services from a total cost of ownership perspective. Manag. Res. Rev. **36**(6), 613–638 (2013)
63. Willcocks, L.P.: Evaluating the outcomes of information systems plans managing information technology evaluation—techniques and processes. In: Strategic Information Management: Challenges and Strategies in Managing Information Systems, pp. 271–294 (2001)
64. Xiong, H., et al.: CloudLightning: a self-organized self-managed heterogeneous cloud. In: Federated Conference on Computer Science and Information Systems (FedCSIS), pp. 749–758 (2017)
65. Yang, H., Tate, M.: A descriptive literature review and classification of cloud computing research. Commun. Assoc. Inf. Syst. **31**(1), 35–60 (2012)